\documentstyle[12pt]{article}

\newcommand{\nc}{\newcommand}
\nc{\beq}{\begin{equation}}
\nc{\eeq}{\end{equation}}
\nc{\beqa}{\begin{eqnarray}}
\nc{\eeqa}{\end{eqnarray}}

\def\ni {\noindent}

\input epsf
\newwrite\ffile\global\newcount\figno \global\figno=1

\def\writedef#1{}
\def\figin{\epsfcheck\figin}\def\figins{\epsfcheck\figins}
\def\epsfcheck{\ifx\epsfbox\UnDeFiNeD
\message{(NO epsf.tex, FIGURES WILL BE IGNORED)}
\gdef\figin##1{\vskip2in}\gdef\figins##1{\hskip.5in}% blank space instead
\else\message{(FIGURES WILL BE INCLUDED)}%
\gdef\figin##1{##1}\gdef\figins##1{##1}\fi}
\def\figinsert{}
\def\ifig#1#2#3{\xdef#1{fig.~\the\figno}
\writedef{#1\leftbracket fig.\noexpand~\the\figno}%
\figinsert\figin{\centerline{#3}}\medskip\centerline{\vbox{\baselineskip12pt
\advance\hsize by -1truein\center\footnotesize{  Fig.~\the\figno.} #2}}
\bigskip\endinsert\global\advance\figno by1}
\def\endinsert{}
\begin{document}

\title{\large{\bf Neutron Star Vortex Dynamics and Magnetic Field
Decay: Implications for High Density Nuclear Matter}}

\author{
Stephen D.H.~Hsu\thanks{hsu@duende.uoregon.edu} \\
Department of Physics, \\
University of Oregon, Eugene OR 97403-5203 \\ \\   }

\date{March, 1999}

\maketitle

\begin{picture}(0,0)(0,0)

\put(350,360){OITS-672}

\end{picture}

\vspace{-24pt}

\begin{abstract}
We investigate the effect of the density-dependent proton and neutron gaps
on vortex dynamics in neutron stars. We argue that the persistence of 
neutron star magnetic fields on timescales of $10^9$ y suggests a
superconducting gap curve with local maximum at intermediate density. 
We discuss the implications for exotic core phenomena such as pion/kaon 
condensation or a transition to quark matter.
 
\end{abstract}

\newpage

In this letter we address the evolution of magnetic fields in neutron stars,
in particular the distribution of magnetic vortices inside the star. 
Residual magnetic fields are believed to persist over very long timescales
($\sim 10^9 y$) in neutron stars.
While naively attributed to the confinement of magnetic flux into vortices
(henceforth, flux vortices or FVs) due to proton superconductivity
\cite{proton}, the phenomena is more involved, and may involve the interaction
of FVs with neutron superfluid vortices (henceforth, SVs) \cite{mag}. 
Here we will argue that a prerequisite for the persistence of magnetic fields, 
as well as for the applicability of models in \cite{mag}, is that the 
proton gap curve $\Delta_p$ have a
certain shape as a function of density within the neutron star. The point is 
that the density-dependent proton gap leads to a force which acts on FVs. At
low densities (in the outer core), this force will always act to eject vortices
into the non-superconducting crust. A simple calculation shows that this proton
gap force dominates any vortex bouyancy effects \cite{buoy}, and leads to 
ejection on timescales of $\sim 10^6$ y. However, if the proton gap decreases 
at higher densities after reaching a local maximum at some intermediate 
density, the sign of the force will reverse and act to anchor vortex segments 
to the core of the neutron star \cite{pethick,jones}. 
We will argue that without 
this effect, interactions between pinned SVs and FVs are 
insufficient to prevent FV ejection.

The phenomenology of magnetic fields in neutron stars has long been of 
interest to those studying pulsar glitches
\cite{glitch}, and has recently been given a prominent role in the magnetar
model of local gamma ray bursters \cite{magnetar}. 
Our main interest here
will be in the dynamics of fluxoids deep within the core of the neutron star,
in particular the forces which act to either anchor or expel them. 
We will conclude with a discussion of the implications of our work on
exotic states of matter in neutron stars.

Below we list some neutron star properties of relevance to our
analysis

\ni $\bullet$ 
Neutron star structure: In the outer layer of thickness $\sim 1$ km,
a lattice of neutron-rich nuclei is surrounded by a neutron superfluid. As
the density increases, conversion of protons and captured electrons into 
neutrons becomes more efficient, and eventually the proton and electron
fraction becomes of order a few percent, sufficient to prevent neutron
decay by Fermi blocking. The neutron superfluid order parameter (see
\cite{HH} for recent computations) is initially in the $^1S_0$ channel,
but probably shifts to the $^3P_2$ channel at higher density, due to
the repulsive core of the neutron-neutron potential. The gap size is of 
order 1 MeV. A proton gap of similar size, leading to superconductivity, 
is also expected in the core region. Due to
uncertainties in the equation of state at high density, the maximum core
density is unknown. Various exotic phenomena such as pion \cite{pion} or
kaon \cite{kaon} condensation, or even a transition to  
quark matter \cite{quark} may occur deep in the core. We note that
in all of these scenarios a superconducting gap which is {\it larger} than
the proton gap is to be expected. (See \cite{colorsc} for recent progress
on the quark color-superconducting gap.)

\ni $\bullet$
Superfluid vortices (SVs) carry the star's angluar momentum in quantized
lines parallel to the spin axis. The have an area density
\beq
n_{SV} ~=~ 2 m_n \Omega / \pi ~\sim~ 10^4 / P {\rm (sec)~~ cm^{-2}}~~.
\eeq
Because of the strong coupling between neutrons and protons, the circulation
of neutrons leads in turn to circulation of protons, and the SVs are 
themselves expected to carry carry magnetic fields.

\ni $\bullet$ 
Magnetic flux vortices (FVs) are the result of the type II superconductivity
in the inner crust and core region. The magnetic field of the star is
confined into individual vortices of flux $\Phi_0 = \pi / e = 2 \cdot 10^{-7}~ 
{\rm Gauss/cm^2}$. The number density of such vortices is
\beq
n_{FV} ~=~ B / \Phi_0 ~\sim~ 10^{19} B_{12} ~{\rm cm^{-2}}~~, 
\eeq
where $B_{12} = B / 10^{12}$ Gauss.
Note that the density of FVs is enormously larger than that of the SVs.

Now let us consider the effect on vortex dynamics of the shapes of
the relevant gap curves. Because the string tension (energy per unit
length) of a vortex behaves as
\beq
\mu = c \Delta^2~~,
\eeq
(where c is a dimensionless constant of order 1)
there is a force per unit length exerted on the vortex due to the variation
of the gap with radial position (density) within the star:
\beq
\vec{f}_{\Delta} ~=~ 2 c \Delta {\partial \Delta \over \partial r} ~\hat{r}~~.
\eeq
The magnitude of this force per unit length is of order
\beq
f_{\Delta} \sim {\rm MeV}^2 / R~~,
\eeq
where R is the characteristic length scale over which the gap varies.
For the FV gap $R \sim R_{NS} \sim 10^4$ m, while for the $^1S_0$
superfluid gap $R \sim 10^3$ m. Comparing with the bouyancy effect of
Muslimov and Tsygan \cite{buoy}, we see that this effect is of similar
but somewhat larger size.

In the region where $\Delta$ is increasing with density, the force will
act to expel vortices. The characteristic time for this to occur depends
on the drag force exerted on the vortex due to interactions with leptons
(at high densities there may be muons present). Since the protons and
neutrons form a superfluid their contribution to the drag is negligible.
The lepton drag force has been considered in some detail by Jones 
\cite{jones}, and is of the order
\beq
f_{drag} \sim {\rm MeV}^3 ~v_{vortex}~~. 
\eeq
Using this result, the terminal velocity can be found and therefore the
expulsion time, which is $\tau_{\Delta} \sim 10^6$ y for magnetic vortices. 
Once a vortex has been expelled into the outer crust, the magnetic field can
decay by ohmic dissipation on timescales of 
$\tau_{\omega} \sim r_c^2 \sigma \sim 10^7$ y, where $r_c$ is the crust
thickness and $\sigma$ the conductivity \cite{ohm}\footnote{Some calculations,
such as that of Sang and Chanmugam \cite{ohm1}, have obtained timescales 
for ohmic decay which are larger than the usual estimate. 
However, it is important to note
that the mechanism described here ejects the magnetic fields into the
{\it outer} crust ($\rho < 10^{12} {\rm g/cm^3}$), where the conductivity
is lower and where even the calculations of \cite{ohm1} yield decay
timescales of order $10^7 y$.}. These timescales are
inconsistent with the observed persistence of magnetic fields of order
$10^{9-10}$ G in millisecond pulsars with ages $10^9$ y. 

In regions where the gap decreases with increasing density, the force
acts to pull the vortex deeper into the star. For example, in the case of
a superfluid vortex, the $^1S_0$ gap falls off after reaching its maximum
at a Fermi momentum $p_F \sim 150$ MeV. In this region an SV is pulled
toward the center of the star, until the sign of the gradient switches again. 
The case of superfluid vortices is complicated, because the superfluid
order parameter switches from $^1S_0$ to $^3P_2$ at $p_F \sim 300 $ MeV.
In addition, because SVs also carry magnetic fields, they are also affected
by the proton gap force gradient. In figure 1 we show the likely
behavior of the neutron and proton gap functions. The leftmost curve shows
the likely behavior of the $^1S_0$ superfluid gap, while the two curves on
the right display possibilities for the superconducting gap. We will refer to
the upper curve, which increases monotonically with density, as curve 1, and
the lower curve as curve 2. Superfluid vortices can
minimize their energy in the region where the superfluid and superconducting
gap curves intersect. The evolution of FVs depends 
crucially on the shape of the $\Delta_p$ curve. If there is no local
maximum (as shown in curve 1), then all FVs will eventually be ejected
from the star. Alternatively, if curve 2 is correct then FVs with 
sufficient length in the attractive core will be anchored against ejection
(see figure 2).
Some sub-population of the FVs could presumably remain indefinitely.

\epsfysize=10 cm
\begin{figure}[htb]
\center{
\leavevmode
\epsfbox{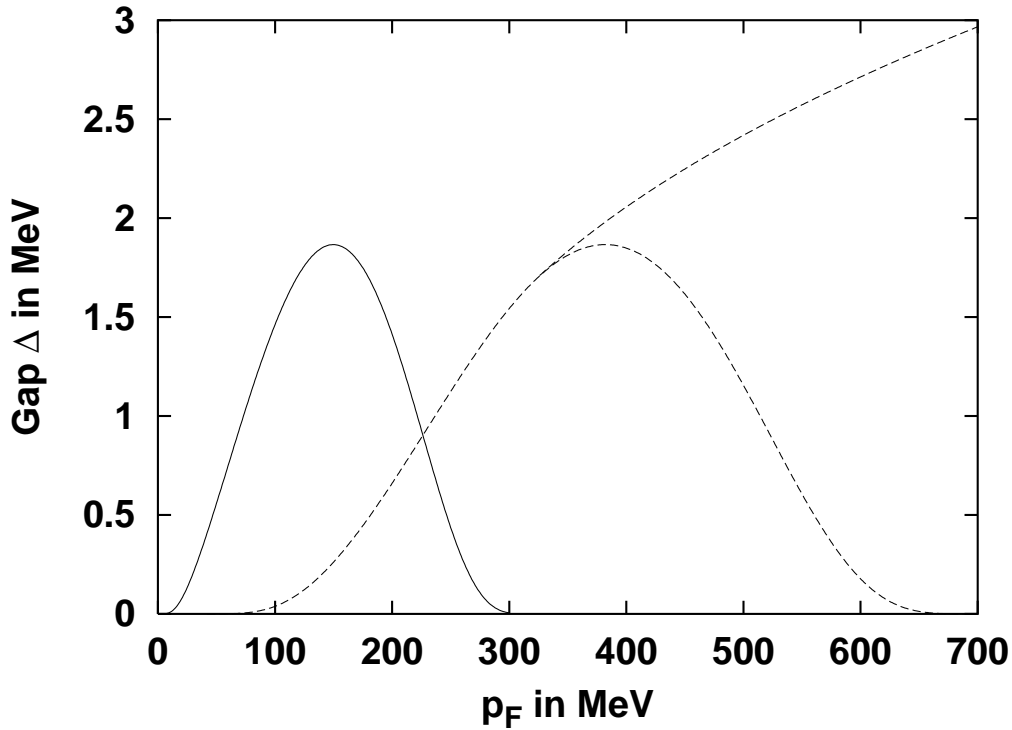}
\caption{Neutron and proton gap curves as a function of neutron
Fermi momentum (all units MeV).} \label{fig1}
}
\end{figure}

One might think that the interactions between FVs and SVs, or their
respective pinning to the crust, might be enough to prevent FV ejection
even in the case of curve 1.
However, since the number of FVs is so much greater than
the number SVs, they will either carry the SVs along in their motion, or
cut through them on their way to the surface. 
(Note that intercommutation of vortices is highly efficient 
\cite{intercommute}, so if a vortex line is cut through it will almost
always reconnect with itself afterwards.) The crustal pinning force on an 
SV is less than of order $\rm MeV^2$, so it is easily overcome by the combined
force exerted by $f_{\Delta}$ through $\sim n_{FV} / n_{SV}$ flux vortices, 
each of order $R_{NS}$ in length. The total force exerted on a single SV
is  
\beq
F_{\Delta} ~\sim~ {n_{FV} \over n_{SV}}~ {\rm MeV^2}~~,
\eeq
which completely dominates any restraining effects on the SV.

The general form of curve 2 in figure 1 is to be expected from standard
calculations, given that pp interactions are attractive at long distances
and have a repulsive core. Of course, medium effects due to the large
density of neutrons will be important and are difficult to account for.
The particular values of curve 2 were obtained using the Fermi surface
effective field theory technique of \cite{HH}, using experimentally
determined pp phase shifts and the beta-stability condition to determine
the proton density relative to the neutron density. 
The result should be accurate at lower densities, but the eventual 
behavior of the curve (i.e. curve 1 vs curve 2) is subject to large
uncertainties. We have argued that the long time persistence of pulsar 
magnetic fields favors case 2. 

As previously mentioned, many of the exotic possibilities for the inner
core behavior (pion or kaon condensation, quark matter) imply
superconducting gaps larger than of order 1 MeV, due to condensation of
electrically charged degrees of freedom: $\pi^\pm$, $K^+$ or a
diquark pair, 
at densities of several times $\rho_0 = 2 \cdot 10^{14} {\rm g/cm^3}$. 
In the case of quark matter \cite{colorsc}, the gap size is expected
to be at least 10 MeV, and perhaps as large as 50-100 MeV.
This would be hard to reconcile with 
curve 2. The transition from normal matter to 
exotic phase would have to occur at sufficiently high density (i.e. at the
far right of figure 1) to allow for a region in the star which remains 
attractive to FVs. The maximum of the proton gap curve in figure 1 is 
already at a density of $\simeq 2 \rho_0$ (and density increases with the
cube of Fermi momentum), so this at most leaves room
for a thin shell of attractive volume. We conclude 
that exotic phases (if they occur at all) 
(1) can only occur at very high density 
($ > ({\rm few}) \rho_0$) and (2) will occupy at most only a small fraction
of the volume of the star.   

\epsfysize=6 cm
\begin{figure}[htb]
\center{
\leavevmode
\epsfbox{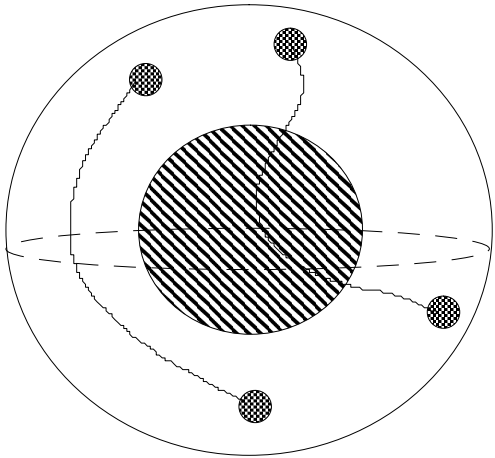}
\caption{Two magnetic vortices, one destined to be expelled, the other
attached to the core.} \label{nstar}
}
\end{figure}

In summary, we have argued that the proton gap curve is likely to exhibit
a local maximum at intermediate density, implying a region 
at higher density which traps flux vortices and disfavoring an exotic
phase at the core.
Vortices which are formed with insufficient 
length in this region will be ejected on timescales of order 
$\tau_{\Delta} \sim 10^6$ y, and 
decay in the outer crust.  As mentioned, 
the asymptotic values of neutron star magnetic
fields are estimated to be less than of order $10^{10}$ G, compared to
$10^{12}$ G or more at formation. It is not known whether the decay of the 
magnetic field is due to accretion or flux decay. If the cause is flux
decay, it would imply that in any (young) neutron star the ejection 
process is under way, with some FVs being pushed into the crust 
at all times. It is not clear what the phenomenological implications
of this are, although the presence of large magnetic fields confined to
the outer crust presumably leads to significant crustal stresses and perhaps
starquake activity. Another issue worth considering is the fate of SVs if
they are carried along in the expulsion of FVs to the surface of the star.
This may lead to spin down which is correlated to the decay of the magnetic
fields. While the causality is different, the phenomenology might resemble
that of models in which magnetic field decay is caused by the flow of SVs
during spin down \cite{glitch}.

\newpage

\bigskip
\noindent 
The author would like to thank Jim Hormuzdiar for discussions.  
This work was supported in part under DOE contracts DE-AC02-ERU3075
and DE-FG06-85ER40224.

%\newpage

\end{document}